\documentclass{article}
\usepackage{dcase2025_techrep}
\usepackage{booktabs}
\usepackage{url}

\title{Learning from Audio-Dependency Errors: Data Curation Strategies Based on Model Confusion Patterns in Audio Question Answering}
\name{Hyeonuk Nam}

\address{Independent Researcher\\
Seoul, South Korea\\
frednam@kaist.ac.kr}

\begin{document}
\maketitle

\begin{abstract}
We frame the system as diagnostic data curation for a large audio-language model: before fine-tuning, we probe Qwen3-Omni-30B-A3B-Instruct under normal, empty-audio, and shuffled-audio conditions to identify how the model's answers change when audio evidence is removed or mismatched. These model confusion patterns are used to bucket training samples into text-prior, shuffle-leak, strong audio-dependent, and hard or misleading cases. Our strongest train-only system fine-tunes only on strong-audio items, where the normal audio-question pair is correct but both counterfactual variants fail, plus a small number of empty-audio negatives and a text-only response normalizer for parse-failed generations. On the official development set, the best train-only system reaches 67.27\% accuracy after response normalization, compared with 65.90\% for our local Qwen3-Omni baseline. Final submissions additionally include models trained using train+development splits and a three-model ensemble.
\end{abstract}

\vspace{-10pt}

\section{Introduction}
\vspace{-10pt}
DCASE 2026 Task 5 evaluates audio-dependent multiple-choice question answering on ADQA-Bench~\cite{dcase2026task5}.
The task differs from generic audio captioning or audio tagging because the model must answer a specific question from the audio, while many choices remain plausible from language priors alone.
In preliminary experiments, we observed that strong open audio-language models often answer a large fraction of samples correctly even when the audio is removed.
This motivated an audio-dependency-aware training recipe inspired by audio contribution analysis in AudioMCQ~\cite{he2026audiomcq}.
Rather than treating all labeled examples as equally useful, we use the base model's errors under counterfactual audio conditions as data-curation signals: samples solved without audio are down-weighted or removed, while samples that fail without the correct audio become the main fine-tuning pool.

\section{Method}
\vspace{-10pt}
\subsection{Base Model, Fine-Tuning, and Inference}
All systems use Qwen3-Omni-30B-A3B-Instruct~\cite{qwen3omni2025,qwen3omnihf}.
We fine-tune with LLaMA-Factory~\cite{zheng2024llamafactory} using 4-bit bitsandbytes quantization and LoRA~\cite{hu2021lora} with rank 4, alpha 8, dropout 0.05, and trainable query and value projections only.
Training uses bfloat16, per-device batch size 8, no gradient accumulation, learning rate $5\times10^{-5}$, cosine scheduling, 3\% warmup, cutoff length 2048, and 2000--3000 maximum update steps depending on the submitted system.
The audio tower and multimodal projector are frozen.
Inference uses greedy decoding with a prompt that asks the model to output only the exact option text.
The training and evaluation code will be released at \url{https://github.com/frednam93/adqa_nam}.

For selected submissions, raw generations are post-processed in two stages.
First, deterministic parsing checks exact option-text matches and simple letter or prefix patterns.
Second, parse-failed responses are passed to a text-only multiple-choice response normalizer.
For single-model submissions we use base Qwen3-Omni as the normalizer; for the three-model ensemble we use Gemma-4-E4B-it so that the total model size remains below the 100B system limit.
The Gemma normalizer is a size-constrained compromise: on the development set it closely matched the base Qwen3-Omni normalizer while allowing us to submit a three-model ensemble within the challenge limit.
The normalizer sees only the candidate choices and the raw generation; it does not receive the audio or the ground-truth label.

\subsection{Prompting and Diagnostic Input Variants}
The default training and inference prompt presents the audio, the question, and four option texts.
The user instruction is to choose the correct option and answer with only the exact option text, rather than a letter.
This exact-option format was chosen because it gave substantially fewer unrecoverable outputs than letter-only or explicit reasoning prompts in our experiments.
Letter-only and brief chain-of-thought prompt variants were also tested and are summarized in Section~\ref{sec:failed}.

For diagnostic analysis, we constructed empty-audio, shuffled-audio, and choice-shuffled variants.
These variants are used only for analysis and data selection, not as direct labels for the final evaluation set.
Table~\ref{tab:diagnostic_scores} shows that the development set remains difficult when the audio is removed or replaced, while a stratified train subset is much easier even under text-only or mismatched-audio conditions.
This mismatch motivated filtering out easy text-prior training examples rather than fine-tuning on the full train split.

\begin{table}[t]
\centering
\small
\caption{Qwen3-Omni diagnostic accuracy under counterfactual inputs.}
\label{tab:diagnostic_scores}
\begin{tabular}{l|rr}
\toprule
Input condition & Dev & Train stratified \\
\midrule
Normal audio-question pair & 65.90 & 89.33 \\
Empty audio & 30.93 & 57.21 \\
Random shuffled audio & 29.25 & 45.70 \\
Same-category shuffled audio & 29.87 & 49.58 \\
Cross-category shuffled audio & 27.44 & 47.27 \\
Choice order shuffled & \textbf{66.27} & 88.85 \\
Audio only, no prompt & 15.00 & 8.61 \\
Generic prompt & 16.24 & 4.36 \\
\bottomrule
\end{tabular}
\vspace{-15pt}
\end{table}

\subsection{Audio-Dependency Buckets}
\label{sec:diagnostics}
Before fine-tuning, we ran diagnostic inference on train and development samples under multiple conditions: normal audio-question pairs, empty-audio questions, and shuffled-audio questions.
For each sample, we parse the model's multiple-choice answer in each condition and record whether it matches the ground-truth answer.
Let $N$, $E$, and $S$ denote correctness under the normal, empty-audio, and shuffled-audio conditions, respectively.
The bucket assignment is based on this correctness pattern, not only on the type of counterfactual input.
We define a \emph{strong audio-dependent} item as $N{=}1,E{=}0,S{=}0$: Qwen3-Omni answers the original audio-question pair correctly, but fails both counterfactual variants.
Items with $N{=}1$ and at least one counterfactual success are treated as text-prior-like, audio-helped, or shuffle-leak cases, while items with $N{=}0$ are separated into hard, misleading, or wrong-normal-but-shuffle-correct candidates.
For the train split, these buckets determine which samples are used for SFT; for the development split, the same diagnostic is reported only for analysis and model selection, and no development or evaluation samples are removed when computing scores.
The goal was not to create a perfect taxonomy, but to separate samples likely solved by text priors from samples where the model depends on the audio.
Table~\ref{tab:bucket_summary} summarizes both the bucket rule and the resulting train/development counts.

\begin{table}[t]
\centering
\small
\caption{Diagnostic audio-dependency buckets. $N$, $E$, and $S$ indicate whether the base model answers correctly with the normal audio-question pair, empty audio, and shuffled audio. Rows are mutually exclusive; * means either value.}
\label{tab:bucket_summary}
\begin{tabular}{l|ccc|rr}
\toprule
Bucket & $N$ & $E$ & $S$ & Train & Dev \\
\midrule
Easy text-prior & 1 & 1 & * & 10648 & 402 \\
Shuffle-leak (audio helped) & 1 & 0 & 1 & 1712 & 283 \\
Strong audio-dependent & 1 & 0 & 0 & 4738 & 374 \\
Hard candidate & 0 & 0 & 0 & 1312 & 318 \\
Misleading or prior-only & 0 & 1 & * & 793 & 95 \\
Wrong but shuffle correct & 0 & 0 & 1 & 277 & 135 \\
\bottomrule
\end{tabular}
\vspace{-15pt}
\end{table}

The train split contains 4,738 strong audio-dependent items, approximately 24.3\% of the full train set.
This full-train filtering pass used 19,480 training items; Qwen3-Omni answered 17,098 correctly in the normal condition, 11,441 with empty audio, and 9,140 with random shuffled audio.
The development set contains 374 strong audio-dependent items, approximately 23.3\%.
High-volume categories include speaker identity, speech content, temporal reasoning, speech paralinguistics, music, and sound events.
This analysis also revealed that many development samples are highly prior-driven; therefore, optimizing only for the aggregate score can favor models that preserve text-prior behavior rather than models that improve audio grounding.

\subsection{Fine-Tuning Data Construction}
\begin{table*}[t]
\centering
\small
\caption{Fine-tuning dataset variants. Counts are training examples after adding negative examples when applicable.}
\label{tab:data_variants}
\begin{tabular}{l|rl}
\toprule
Variant & Count & Description \\
\midrule
Strong & 4,738 & Normal correct; empty/shuffled fail \\
Strong+hard & 6,050 & Strong + hard \\
Non-easy & 7,762 & Strong + hard + shuffle-leak \\
Strong+empty 2.5\% & 4,856 & Strong + 2.5\% empty-audio unknowns \\
Strong+empty 5\% & 4,975 & Strong + 5\% empty-audio unknowns \\
Strong+empty 7.5\% & 5,093 & Strong + 7.5\% empty-audio unknowns \\
Strong+empty 10\% & 5,212 & Strong + 10\% empty-audio unknowns \\
Strong+empty 20\% & 5,686 & Strong + 20\% empty-audio unknowns \\
Strong+shuffle 5\% & 4,975 & Strong + shuffled-audio unknowns \\
Strong+empty 5\%+shuffle 5\% & 5,212 & Strong + 5\% empty + 5\% shuffled \\
Category-balanced & 10,594 & Category-resampled broad pool + 5\% empty \\
\bottomrule
\end{tabular}
\vspace{-15pt}
\end{table*}

The main supervised fine-tuning set contains only strong audio-dependent training items selected by the three-run diagnostic filter.
This conservative rule removes easy text-prior samples and excludes normal-wrong hard samples from the primary recipe because their targets may require capabilities the current base model does not reliably expose.
We also tested broader subsets that include hard audio-dependent candidates, weak/non-easy audio-dependent items, and shuffle-leak items, but these did not improve development accuracy.
Here, \emph{Strong+hard} adds the hard audio-dependent bucket to the strong bucket, and \emph{Non-easy} adds both hard and shuffle-leak buckets.

To discourage unsupported guessing, we add empty-audio negative examples.
For these examples, the audio input is removed and the target response is ``Cannot be determined from the audio.''
The system message is also modified to explicitly allow this unknown response when the audio is missing or insufficient.
We tested several negative-example ratios, from 2.5\% to 20\% of the positive set used by each variant.
For the strong-audio variants, these percentages are measured relative to the 4,738 strong positive examples; for example, 5\% adds 237 unknown-target negatives.
We additionally tested shuffled-audio unknown examples, where the question is paired with unrelated audio and the target is again the unknown response.
The combined empty+shuffle 10\% setting adds 5\% empty-audio negatives and 5\% shuffled-audio negatives.
Finally, the category-balanced broad pool starts from the broad non-easy-style pool with a small easy-prior component, resamples it to make the ten main question categories roughly uniform, and adds 5\% empty-audio unknowns.
We sweep the negative-example ratio and checkpoint step, using the development set for model selection.
Table~\ref{tab:data_variants} summarizes the main fine-tuning datasets.

\section{Development Results}
\begin{table}[t]
\centering
\small
\caption{Main train-only development results.}
\label{tab:main_results}
\begin{tabular}{l|rrr}
\toprule
System & Step & Strict & Judge \\
\midrule
Qwen3-Omni baseline & -- & 65.90 & -- \\
Strong only & 3000 & 64.97 & 66.21 \\
Strong + empty 2.5\% & 3000 & 65.77 & 67.02 \\
Strong + empty 5\% & 1000 & 64.78 & 66.09 \\
Strong + empty 5\% & 2000 & \textbf{66.02} & \textbf{67.27} \\
Strong + empty 5\% & 3000 & 65.28 & 66.52 \\
Strong + empty 7.5\% & 2500 & 65.34 & 66.52 \\
Strong + empty 10\% & 1000 & 64.90 & 66.15 \\
Strong + empty 20\% & 3000 & 65.09 & 66.21 \\
Strong + empty 5\% + shuffle 5\% & 1000 & 64.97 & 66.27 \\
\bottomrule
\end{tabular}
\vspace{-15pt}
\end{table}
\vspace{-10pt}
Table~\ref{tab:main_results} summarizes the main train-only systems.
The judge column applies the Qwen3-Omni text-only response normalizer only to parse-failed predictions.
The best train-only setting is the 5\% empty-audio recipe at 2000 steps.
The 2.5\% recipe at 3000 steps is the second strongest setting after response normalization.
Higher empty ratios and combined empty+shuffle negatives did not outperform the best two settings.
Letter-only and explicit chain-of-thought inference prompts were also worse than exact-option generation: for the best checkpoint, letter-only reached 60.17\% after normalization and CoT-then-letter reached 65.34\%, both below the default prompt.
The exact-option prompt sometimes produces long non-option text, but the text-only normalizer recovers most such parse failures without changing already parsed predictions.

Table~\ref{tab:subset_results} gives a more complete view of the dataset-selection sweep.
The most important negative result is that simply adding more difficult or weakly audio-dependent samples degrades accuracy.
The best result came from the smaller strong-audio subset with a small empty-audio negative set, not from the larger non-easy or category-balanced pools.

\begin{table}[t]
\centering
\small
\caption{Additional Qwen3-Omni SFT ablations.}
\label{tab:subset_results}
\begin{tabular}{l|rrr}
\toprule
Training variant & Step & Strict & Judge \\
\midrule
Strong & 1000 & 63.91 & 65.28 \\
Strong & 2000 & 64.59 & 65.84 \\
Strong & 3000 & 64.97 & 66.21 \\
Strong+hard & 1000 & 61.36 & 62.54 \\
Strong+hard & 3000 & 61.11 & 62.35 \\
Non-easy & 1000 & 61.85 & 62.97 \\
Non-easy & 3000 & 60.55 & 61.61 \\
Non-easy+empty 5\% & 1000 & 62.91 & 64.22 \\
Non-easy+empty 5\% & 3000 & 62.35 & 63.53 \\
Non-easy+shuffle 5\% & 2000 & 61.92 & 63.16 \\
Non-easy+empty 5\%+shuffle 5\% & 3000 & 62.60 & 63.91 \\
Strong+shuffle 5\% & 3000 & 64.41 & 65.65 \\
Strong+empty 5\%+shuffle 5\% & 1000 & 64.97 & 66.27 \\
Strong+empty 2.5\% & 2000 & 65.46 & 66.83 \\
Strong+empty 2.5\% & 3000 & 65.77 & 67.02 \\
Strong+empty 5\% & 2000 & \textbf{66.02} & \textbf{67.27} \\
Strong+empty 7.5\% & 2500 & 65.34 & 66.52 \\
Strong+empty 20\% & 3000 & 65.09 & 66.21 \\
\bottomrule
\end{tabular}
\vspace{-15pt}
\end{table}

\subsection{Additional Ablations and Failed Directions}
\label{sec:failed}
\begin{table}[t]
\centering
\small
\caption{Representative failed or deprioritized experiments on the development set.}
\label{tab:failed}
\begin{tabular}{lrr}
\toprule
Experiment & Strict & Judge \\
\midrule
FunAudioChat~\cite{funaudiochat2025,funaudiollm2024} base & 54.26 & -- \\
AudioFlamingo3~\cite{goel2025audioflamingo3} rerun & 54.32 & -- \\
Answer-only Qwen SFT, 9k & 62.85 & -- \\
Silent CoT SFT, 3k & 61.05 & -- \\
Explicit CoT SFT, 3k & 58.49 & -- \\
Category-balanced & 64.09 & -- \\
Strong+10\% cat-balanced mix & 65.21 & -- \\
Empty5 repeat, seed2 best & 64.16 & -- \\
Empty5 repeat, seed3 best & 65.21 & -- \\
SimPO~\cite{meng2024simpo} continuation, 500 & \textbf{65.65} & \textbf{66.96} \\
DPO~\cite{rafailov2023dpo} continuation, 500 & 65.15 & 66.46 \\
SFT-init GRPO~\cite{shao2024deepseekmath}, 100 & 65.09 & 66.21 \\
SFT-init DAPO-lite~\cite{yu2025dapo}, 100 & 65.21 & 66.40 \\
Empty-aware GRPO~\cite{shao2024deepseekmath}, 100 & 65.09 & 66.46 \\
Empty-aware DAPO-lite~\cite{yu2025dapo}, 100 & 64.72 & 65.90 \\
\bottomrule
\end{tabular}
\vspace{-15pt}
\end{table}

We evaluated several alternative improvement directions; none replaced the simple strong-audio SFT recipe.
Table~\ref{tab:failed} lists the main negative results.
The early answer-only Qwen3 fine-tuning runs degraded the base model, even though the base model was already the strongest single model among the tested backbones.
Category-level model routing provided only a small oracle gain: choosing the best model per rule-based category would reach 66.83\%, compared with 65.90\% for Qwen3-Omni base.
The much higher per-sample oracle of 90.85\% indicates complementary errors, but coarse category routing alone is insufficient.
We also tested full category rebalancing and additive category-balanced mixing.
The category-balanced setting replaces the strong-only positives with the resampled broad pool in Table~\ref{tab:data_variants}.
The additive mix keeps the strong-empty5 recipe as the main pool but samples 10\% of training rows from the category-balanced pool.
The best full rebalanced run reached 64.09\% strict accuracy, and the best additive variant reached 65.21\%.
Both remained below the strong-empty recipe.
Repeating the 5\% empty-audio recipe with two additional random seeds also failed to improve the original run: the best strict accuracies were 64.16\% and 65.21\%.
Preference/RL continuation also failed to exceed the 5\% empty-audio SFT checkpoint.
The tested preference objectives were DPO~\cite{rafailov2023dpo}, SimPO~\cite{meng2024simpo}, ORPO~\cite{hong2024orpo}, and KTO~\cite{ethayarajh2024kto}; the RL-style objectives were GRPO~\cite{shao2024deepseekmath} and a small DAPO-style variant~\cite{yu2025dapo}.
ORPO collapsed into invalid outputs in our run, and KTO was not used because the audio metadata path was dropped by the training pipeline.

\section{Discussion}
\vspace{-10pt}
\subsection{Error Analysis}
Comparing the best train-only system with the Qwen3-Omni baseline shows a clear trade-off.
After response normalization, the fine-tuned model fixes 130 baseline errors but introduces 108 regressions.
Many gains are in speech, voice, and phonetics, music, counting, and sound-event questions.
Regressions are also concentrated in speech, voice, and phonetics and music, suggesting that fine-tuning improves audio dependence but can damage some prior knowledge or fine-grained perception.
In the bucket analysis, the fine-tuned model improves hard-all-wrong and misleading-prior buckets, but loses accuracy on text-prior-easy and shuffle-leak buckets.
This supports our final choice to include one train-only system and two train+development variants rather than relying on a single heavily specialized model.

A notable failure mode is that the fine-tuned model can become more conservative about prior-based answers while still not improving the underlying acoustic perception enough for every category.
For example, it improves several hard samples that require counting speakers, identifying subtle voice attributes, or distinguishing sound events, but it can regress on samples where the base model already used a robust prior or memorized common answer pattern.
Thus, our submitted systems intentionally preserve diversity across the 5\% and 2.5\% empty-negative recipes instead of selecting only one checkpoint.

\subsection{Submitted Systems}
We prepare four candidate submissions:
\begin{itemize}
    \item \textbf{System 1}: train-only strong-audio SFT with 5\% empty-audio negatives at 2000 steps, followed by base Qwen3 response normalization.
    \item \textbf{System 2}: train+development strong-audio SFT with 5\% empty-audio negatives at 2000 steps, followed by base Qwen3 response normalization.
    \item \textbf{System 3}: train+development strong-audio SFT with 2.5\% empty-audio negatives at 3000 steps, followed by base Qwen3 response normalization.
    \item \textbf{System 4}: an ensemble of Systems 1, 2, and 3. Parse-failed outputs are normalized by Gemma-4-E4B-it, then the three systems vote. If all three systems disagree, System 2 is used as the tie breaker.
\end{itemize}
System 4 is a diversity-oriented candidate rather than a confirmed improvement: its vote may reduce model-specific errors, but its Gemma-4-E4B-it normalizer keeps the ensemble within the challenge size limit.
Systems 2 and 3 combine 4,738 train strong-audio items with 374 development strong-audio items; after empty-audio negatives, they contain 5,368 and 5,240 SFT examples, respectively.
Because the development labels are used in Systems 2--4, no held-out development score is reported for those final variants.

\section{Conclusion}
\vspace{-10pt}
The most effective intervention is not a more complex backbone or RL objective, but data curation based on how the model fails when audio evidence is removed or mismatched.
Careful removal of easy text-prior samples and a small amount of empty-audio negative training provide modest but consistent gains, suggesting that learning from audio-dependency errors is a practical way to adapt large audio-language models for ADQA-style evaluation.

\bibliographystyle{IEEEtran}
\bibliography{references}

\end{document}